\documentclass{webofc}
\usepackage[varg]{txfonts}   
\usepackage{slashed}
\begin{document}
\title{Electromagnetic transition form factors of baryons\\ 
in a relativistic Faddeev approach}
%
%

\author{\firstname{Reinhard} \lastname{Alkofer}\inst{1}\fnsep\thanks{\email{reinhard.alkofer@uni-graz.at}} \and
        \firstname{Christian S.} \lastname{Fischer}\inst{2} \and
        \firstname{H\`elios} \lastname{Sanchis-Alepuz}\inst{1}
}
\institute{Institut f\"ur Physik, Karl-Franzens--Universit\"at Graz, NAWI Graz, 8010 Graz, Austria\\
\and
Institut f\"ur Theoretische Physik, Justus-Liebig--Universit\"at Giessen, 35392 Giessen, Germany        
}

\abstract{
The covariant Faddeev approach which describes baryons as relativistic three-quark bound states  
and is based on the Dyson-Schwinger and Bethe-Salpeter equations of QCD is briefly reviewed. 
All elements, including especially the baryons' three-body-wave-functions, the quark propagators 
and the dressed quark-photon vertex, are calculated from a well-established approximation for 
the quark-gluon interaction.
Selected previous results of this approach for the spectrum and elastic electromagnetic form
factors of ground-state baryons and resonances are reported. 
The main focus of this talk is a presentation and discussion of results 
from a recent investigation of the electromagnetic transition form factors between 
ground-state octet and decuplet baryons as well as the octet-only $\Sigma^0$ to $\Lambda$ transition. 
}
\maketitle
\section{Introduction}
\label{intro}

Over the last decades we learned that the structure of baryons, including the nucleon, 
shows many previously unexpected features. Being composites of three valence quarks, 
a quark-antiquark sea, and glue, 
baryons are surprisingly complicated objects. Although there is no doubt that a description of
baryons is possible within Quantum Chromo Dynamics (QCD) corresponding investigations are very
challenging and computationally demanding. A prominent example is 
hadron spectroscopy from lattice QCD calculations: It had 
to overcome many technical obstacles before it reached today's precision, see, {\it e.g.},
Daniel Mohler's talk at this conference \cite{Mohler}. 

From those and many other investigations it became evident that the internal structure of 
mesons and baryons is very sensitive to the way quarks and gluons interact at sub-GeV scales. 
Even more than confinement the dynamical breaking of chiral symmetry is an important phenomenon
shaping the arrangement of the hadrons' constituents. Therefore very likely only the interplay of 
different theoretical techniques and their confrontation with the available experimental data will 
allow for an  understanding of the intricacies of the hadrons' structure.

In particular, the hadrons' electromagnetic form factors provide information on how the electric 
charges and electromagnetic multipole moments are distributed. In the spacelike
momentum region they carry information about the hadrons' electromagnetic interaction 
with other charged particles via the exchange of virtual photons.  In addition, 
transition form factors reveal further properties. An important example is the 
electromagnetic nucleon to Delta transition for which the experimental results unequivocally 
demonstrate the existence of deformations. As will be detailed below these deviations 
from sphericity have a straightforward explanation from the properties of relativistic 
fermions, see \cite{Sanchis-Alepuz:2017mir,Eichmann:2016yit} and references therein.
This example makes clear that the objective of the approach presented in this talk 
is much more to gain qualitative insight into the physics governing the structure and dynamics
of hadrons than to produce precise numbers for hadronic observables.

The main idea of the here presented  relativistic Faddeev approach 
\cite{Eichmann:2009zx,Eichmann:2009qa} is to employ the gained
knowledge on the QCD Green functions in the Landau gauge and to use the fact that baryons will 
appear as poles in the six-quark Green funcion. In a similar way as done for the Bethe-Salpeter
equations in the relativistic two-body problem (see, {\it e.g.}, Chapter 6 of 
ref.\ \cite{Alkofer:2000wg}) one obtains then fully Poincar\'e-covariant bound-state equations.
It should be emphasized here that there is nothing special about extracting bound state properties
in quantum theories from the singularities of amplitudes. On the contrary, the textbook treatment
of two-body problems in Quantum Mechanics by mapping the two-body Schr\"odinger equation to the one
for the relative motion and a central potential is the highly exceptional case. But also in 
Schr\"odinger theory the scattering amplitudes obtained from such potentials display poles in the
complex energy plane at $iE_{B,n}$, {\it i.e.}, at the bound state energies but on the imaginary axis.
It is highly instructive to trace the origin of these poles which is possible already at the level
of the one-dimensional Schr\"odinger equation and the simple potential well: One requires a
non-vanishing wave function without an incident wave which leads then to the quantized energy 
eigenvalue condition $E=   iE_{B,n}$, see, {\it e.g.}, Sect.\ 3.7.1 of ref.\ \cite{SchwablQM}.

The solution of the Poincar\'e-covariant Faddeev equation provides besides the bound-state masses
for baryons also the related amplitudes. These serve then in turn as input for the calculation 
of baryon obervables. In this talk we will focus on baryon form factors, and hereby especially 
on the transition form factors. The numerical techniques necessary for this kind of calculations
--- solving first Dyson-Schwinger equations for the Green functions and second Bethe-Salpeter type
 equations for the bound-state masses and amplitudes, calculating third from them the integrals 
for the form factors --- are quite involved, and correspondingly over the last decades many detailed
descriptions of such procedures have been published. For these technical issues we refer here to the
most recent publication on these practical matters, ref.\  \cite{Sanchis-Alepuz:2017jjd},
as well as references therein.

\section{Relativistic Three-Fermion Bound State Equations}
\label{sec2}

\begin{figure}[t]
\begin{center}
\includegraphics[width=0.7\textwidth]{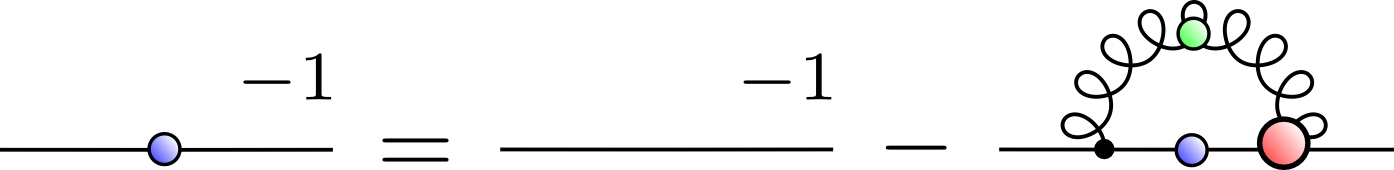}
\caption{The DSE for the quark propagator.}
\end{center}
\label{QDSE}
\end{figure}

As stated in the introduction the here presented Faddeev approach is based on the 
QCD Green functions in the Landau gauge. As valence quarks are those constituents of hadrons
which determine their quantum numbers the most important such function for the 
description of mesons and baryons within bound state equations is the quark 
two-point function, respectively, the quark propagator $S$. Its dressing by
gluonic interactions can be expressed via its Dyson--Schwinger equation (DSE)
\begin{align}\label{eqn:quarkDSE}
 S^{-1}(p)=S^{-1}_0(p)+Z_{1f}\int \frac{d^4q}{(2\pi)^4} \gamma^\mu
           D_{\mu\nu}(p-q)\,S(q)\,\Gamma^\nu_{gqq}(p,q)\, ,
\end{align}
see fig.\ \ref{QDSE} for a pictorial presentation.
Hereby, $S_0$ denotes the renormalized tree-level propagator 
$S^{-1}_0(p)=Z_2\left(i\slashed{p}+m\right)$, and $Z_2$ as well as $Z_{1f}$
are renormalisation constants. $m$ is the bare or current quark mass which, for the 
different flavours, will be provided as a parameter. (Throughout this 
study we employ the isospin limit, and set up and down current quark mass to an equal value.)

The solution of this DSE necessitates in principle knowledge about  the full quark-gluon 
vertex $\Gamma^\nu_{gqq}$ and the gluon propagator $D^{\mu\nu}$. Although calculations 
including quite accurate results for both these functions in the determination of the 
properties of light mesons have been successfully performed \cite{Williams:2015cvx} 
we use here a generalized rainbow truncation and approximate the quark-gluon vertex
by its bare form multiplied with a suitably adjusted function of the gluon momentum
$k=p-q$,  $\Gamma^\nu_{gqq}(p,q) = \gamma^\nu f(k^2)$. The requirement to preserve 
chiral symmetry leads then to a form of the quark-quark kernel,
\begin{align}\label{eqn:kernelRL}
K^{\mathrm{2-body}} = \left[\gamma^\mu \otimes \gamma^\nu\right] f(k^2) D_{\mu\nu}(k)\, ,
\end{align}
which is considerably easier to implement than in a higher-order chiral-symmetry preserving
truncation, see, {\it e.g.}, ref.\ \cite{Williams:2015cvx}. 

\begin{figure}[!b]
\begin{center}
\includegraphics[width=\textwidth]{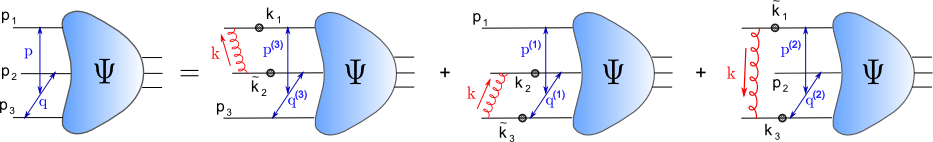}
\caption{The covariant Faddeev equation in rainbow-ladder approximation.}
\end{center}
\end{figure}

In order to make structures transparent and not to clutter all expressions with their many 
different indices one needs in an actual calculation we will present the content of the 
relativistic three-fermion bound state equation in a compact matrix notation. Besides 
summation over discrete indices it includes also integration over repeated  continuous variables.
As the three-particle-irreducible interaction within three quarks seems to be of minor 
importance \cite{Eichmann:2016yit} we are employing the Faddeev approximation and keep
only the two-particle-irreducible interaction. The Faddeev equation is then the permuted sum 
of two-body quark-quark kernels $K^{\mathrm{2-body}}$~\cite{Eichmann:2009qa,SanchisAlepuz:2011jn}:
\begin{align}\label{eqn:compactBSE}
\Psi = \sum_{a=1}^3 \big[K^{\mathrm{2-body}}\big]_{(a)}\,G_0\,\Psi\,.
\end{align} 
Hereby, the index $a$ determines the interacting pair by labeling the spectator quark.
The function $G_0$ describes the free propagation of three quarks and is given by the
disconnected product of three full quark propagators $S(p_i)$.

The interaction of a three-quark system with a single photon is described by the conserved 
current
\begin{align}\label{eq:current_general}
 J^\mu=&~\bar{\Psi}_f\left(G^{\mu}_0 - G_0K^\mu G_0 \right)\Psi_i\,,
\end{align}
where the incoming and outgoing baryon is represented by the Faddeev
amplitudes $\Psi_i$ and $\Psi_f$, respectively.  $G^\mu_0$ denotes the impulse-approximation diagrams
\begin{align}\label{eq:gauged_G0}
 G_0^\mu=&~\left(S_1\,\Gamma^\mu S_1\right)S_2\,S_3+S_1\left(S_2\,\Gamma^\mu
                 S_2\right)S_3+S_1\,S_2\left(S_3\,\Gamma^\mu S_3\right) ,
\end{align}
where $\Gamma^\mu$ is the fully dressed quark-photon vertex. $K^\mu$ stands for the interaction of
the photon with the Faddeev kernel. For this quantity the only diagrams that survive in 
rainbow-ladder approximation are those
where the two-body kernel is a spectator:
\begin{align}\label{eq:gauged_K}
 K^\mu=&~ \Gamma^\mu_1\,K_{23} + \Gamma^\mu_2\,K_{31} + \Gamma^\mu_3\,K_{12}\,.
\end{align}

For consistency it is important that the quark-photon vertex is calculated from its 
Dyson-Schwinger equation~\cite{Maris:1999bh}. 
The Faddeev amplitudes, including their full Dirac--flavor structure, are determined by solving 
the three-body Faddeev equation with the interaction kernel $K$. The numerical solution of this
integral equations rests hereby on the use of pseudo-spectral methods, especially on multiple
expansions in Chebychev polynomials, see, {\it e.g.}, \cite{Sanchis-Alepuz:2017jjd}.

\section{Structure of Baryonic Bound State Amplitudes}

As this will become important in the interpretation of the results we summarise briefly 
the structure of the baryonic bound state amplitudes. Coupling three quarks which are 
spin-1/2 Dirac fermions together in a Poincar\'e covariant way as well as requiring baryons to
be colour singlets and the wanted flavour state determines the elements of a tensorial 
decomposition of baryons' amplitudes \cite{Carimalo:1992ia,Eichmann:2016yit}. 
Phrased otherwise:  The structure of the baryonic bound state amplitudes is 
{\em independent of any truncation of the bound state equation,}  only Poincar\'e covariance 
and parity invariance are exploited to derive them. They include all possible internal spin 
and orbital angular momenta of the considered baryon.
For a positive-parity, positive-energy (particle) baryon such an amplitude consists of
in a partial-wave decomosition in the baryon's rest frame of 
8 $s$-wave, 36 $p$-wave and 20 $d$-wave for spin-1/2 baryons \cite{Eichmann:2009qa}
and 4 $s$-wave, 36 $p$-wave, 60 $d$-wave amplitudes and 28 $f$-wave amplitudes for spin-3/2 
\cite{SanchisAlepuz:2011jn,Nicmorus:2010sd}. Their relative importance are, of course, determined
by the underlying interaction kernel. Although the results presented below indicate that for positive-parity ground-state
baryons an $s$-wave ampitude is the single strongest one typically all others together are of equal 
importance in practically most observables.

An important exception is given if one considers the deviaton from sphericity for a given baryon. 
Here, the $p$-wave amplitudes play a significant role. For a given $s$-wave contribution which is 
related to two upper spinor entries of the baryon's four-spinor the $p$-wave is located in the lower 
half of the spinor. Those spinors are in the context of Dirac spinors related to anti-fermions,
and thus their existence is purely due to the nature of relativistic quantum fields. Therefore, we conclude 
already at this point that the Poincar\'e covariant Faddeev approach offers a very natural explanation
why the nucleon is not spherically symmetric: It is first a relativistic effect, and it is second a 
quantum effect due to existence of anti-particles in a relativistically invariant quantum field theory.

Of course, such an explanation does not exclude that dynamically generated $d$-waves also contribute
to the deviation from sphericity, and as a matter of fact, in the results presented below they do. 
However, the difference is on a qualitative level. Only a fully relativistic approach necessitates to
describe baryons and especially the nucleons as spinors being related to a representation of a 
Clifford algebra. On the other hand, in such an approach such contributions are unavoidable. This leads
to the conclusion that in fully relativistic approach based on binding three Dirac-fermionic quarks 
together spherical baryons are virtually impossible, to obtain them one need an incredible amount of 
fine-tuning not only in the strength of the interquark interaction but also in its tensorial structure.

\section{Earlier selected Results}

\begin{figure}[b]
\begin{center}
\includegraphics[width=0.9\textwidth]{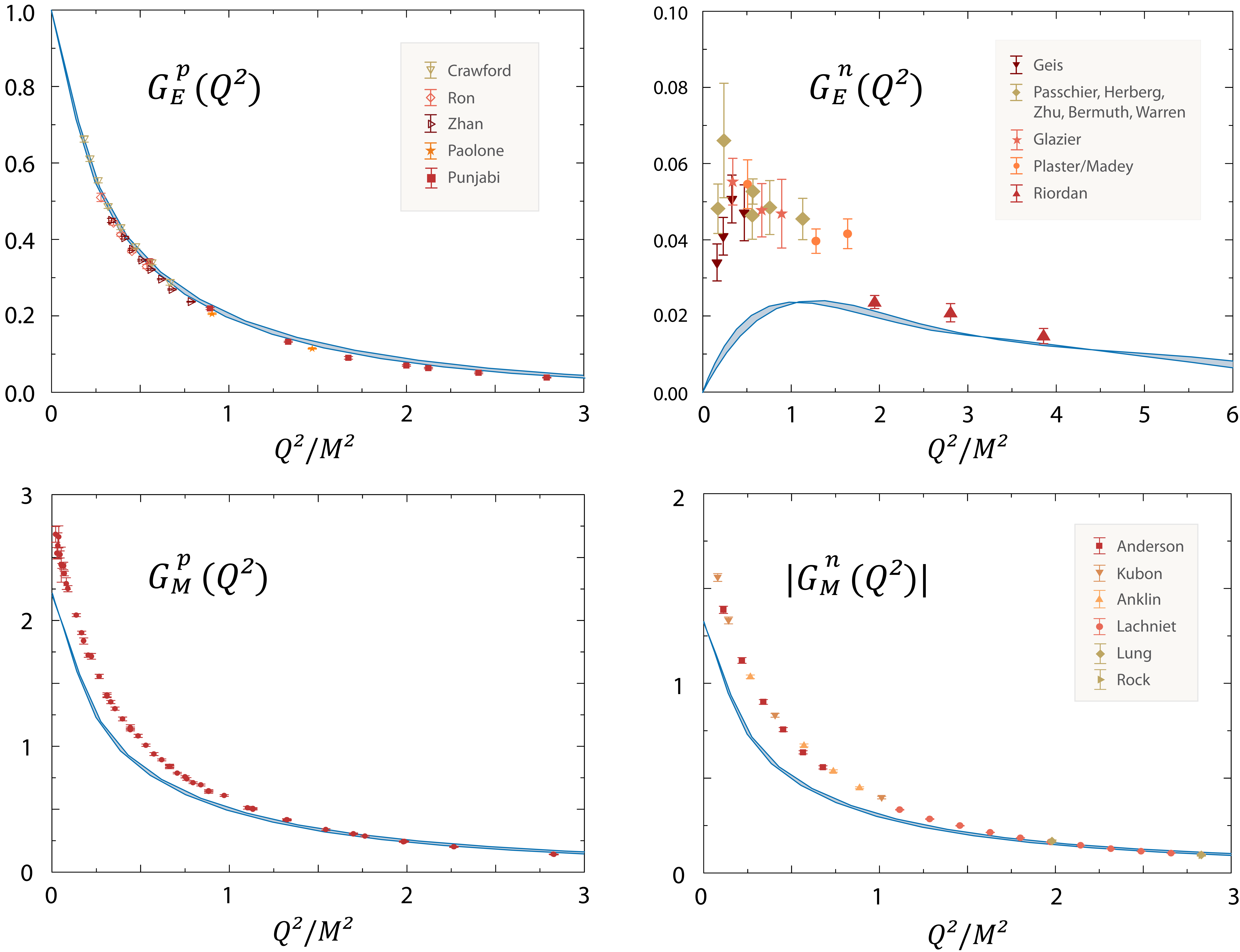}
\caption{Nucleon electromagnetic form factors $G_E$ and $G_M$ for the proton (left panels) and neutron (right panels) compared to experimental data; see~\cite{Eichmann:2011vu} for references.} \label{fig:nucleon-ffs}
\end{center}
\end{figure}

\begin{figure}[t]
\begin{center}
\includegraphics[width=0.9\textwidth]{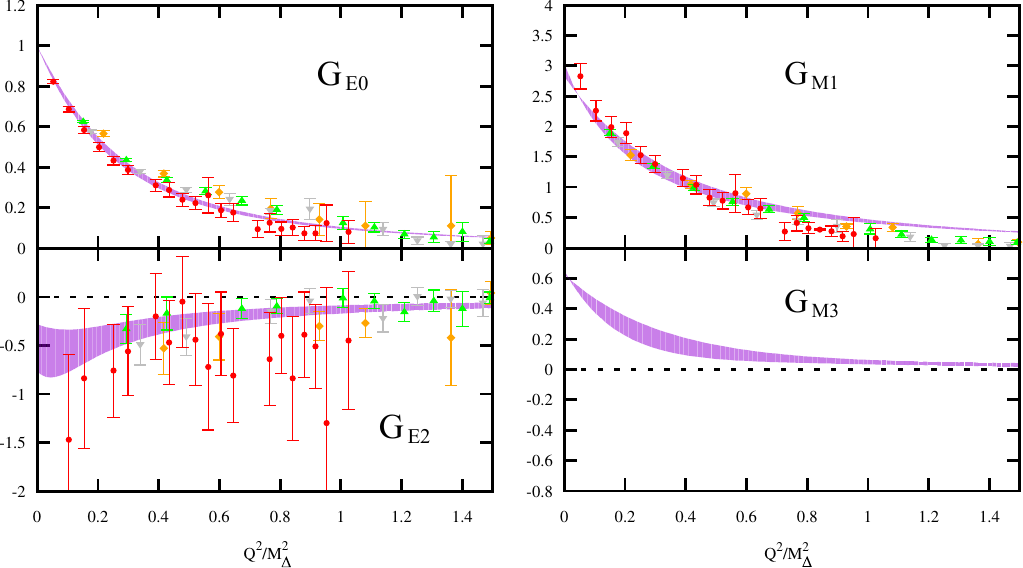}
\caption{$\Delta$ electromagnetic form factors compared to lattice results from \cite{Alexandrou:2009hs}; see~\cite{Sanchis-Alepuz:2013iia} for references.} \label{fig:delta-ffs}
\end{center}
\end{figure}

The electromagnetic structure of baryons is given by several dimensionless and 
Lorentz-invariant form factors.
Hereby, the nucleon as spin-1/2 baryon possesses two form factors which we choose in 
our calcuation as the two Sachs form factors $G_E$ and $G_M$.
The $\Delta$ electromagnetic  are $G_{E0}$, $G_{M1}$, $G_{E2}$ and $G_{M3}$: electric monopole, magnetic dipole, electric quadrupole and magnetic octupole.
Previously obtained results for the $N$ and $\Delta$ electromagnetic form factors  are 
shown in Figs.~\ref{fig:nucleon-ffs} and~\ref{fig:delta-ffs}; details of the respective calculations can be found in Refs.~\cite{Eichmann:2011vu,Sanchis-Alepuz:2013iia}; see also~\cite{Eichmann:2011pv} for results on axial form factors.


  In all cases the comparison with experimental data and lattice results performs rather well.
Note that the only model input in all calculations is the quark-gluon interaction in Eq.~\eqref{eqn:kernelRL}, and therefore this agreement is quite remarkable.
This can be attributed to some extent to the global symmetries of QCD, especially its pattern of 
spontaneous chiral symmetry breaking, which are respected in every step of these calculations.
Furthermore, charge conservation at $Q^2=0$ is not imposed but a consequence of the 
underlying Ward-Takahashi identity which is kept if the quark-photon vertex is calculated consistently.
Note, however, that obtaining the quark-photon vertex from a Dyson-Schwinger equation includes 
automatically the contribution of the vector ($\rho$) meson as the latter as quark-antiquark bound state 
contributes to this vertex.

Nevertheless, the neglect of dynamical pions (as Goldstone bosons of spontaneously broken
chiral symmetry) in the kernel of the Faddeev equation implies visible defects in the form factors
at low $Q^2$, especially in the electric neutron form factor.
Summarizing recent investigations on this topic, see,{\it e.g.}, ref.\ \cite{Eichmann:2016yit} and 
references therein, these deficiences are symptoms of missing meson-cloud effects, which are known to
enhance magnetic moments and charge radii,  especially for small pion masses as the physical one.
However, the rainbow-ladder truncation employed here does not include such dynamical pion effects, and 
especially not decays;
it produces stable bound states that do not decay and can be viewed as the `quark core' contribution 
that is stripped from its pion cloud  \cite{Eichmann:2007nn}.

\section{Transition form factors }

\subsection*{$N\to\Delta\gamma$ transition}

The $N\to\Delta\gamma$ transition form factors are usually discussed in terms of the 
magnetic dipole $G_M^\ast$ and the electric and Coulomb quadrupole ratios $R_{EM}$ and $R_{SM}$:
\begin{flalign}
R_{EM} = -\frac{G^*_{E}}{G^*_M}, \hspace*{10mm} R_{SM}=-\frac{M_N^2}{M_\Delta^2}\sqrt{\lambda_+\lambda_-}\frac{G^*_C}{G^*_M},
\end{flalign}
which at zero momentum are sensitive to the quadrupole moments of the transition.

\begin{figure}[t!]
\begin{center}
\resizebox{0.55\textwidth}{!}{\includegraphics{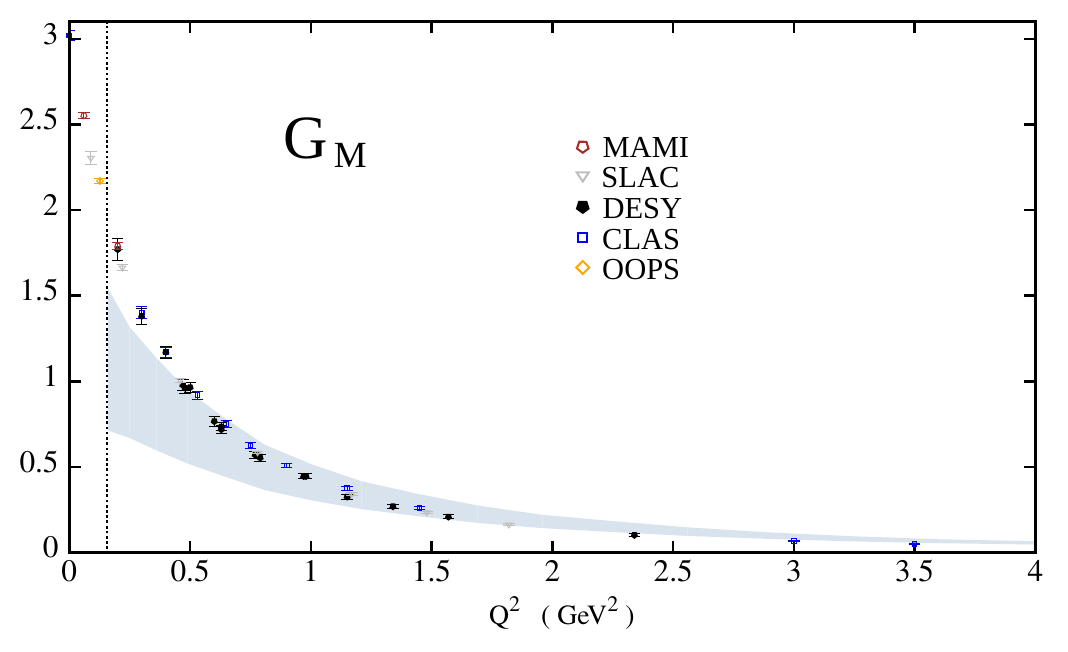}}
\end{center}
\begin{center}
\resizebox{0.45\textwidth}{!}{\includegraphics{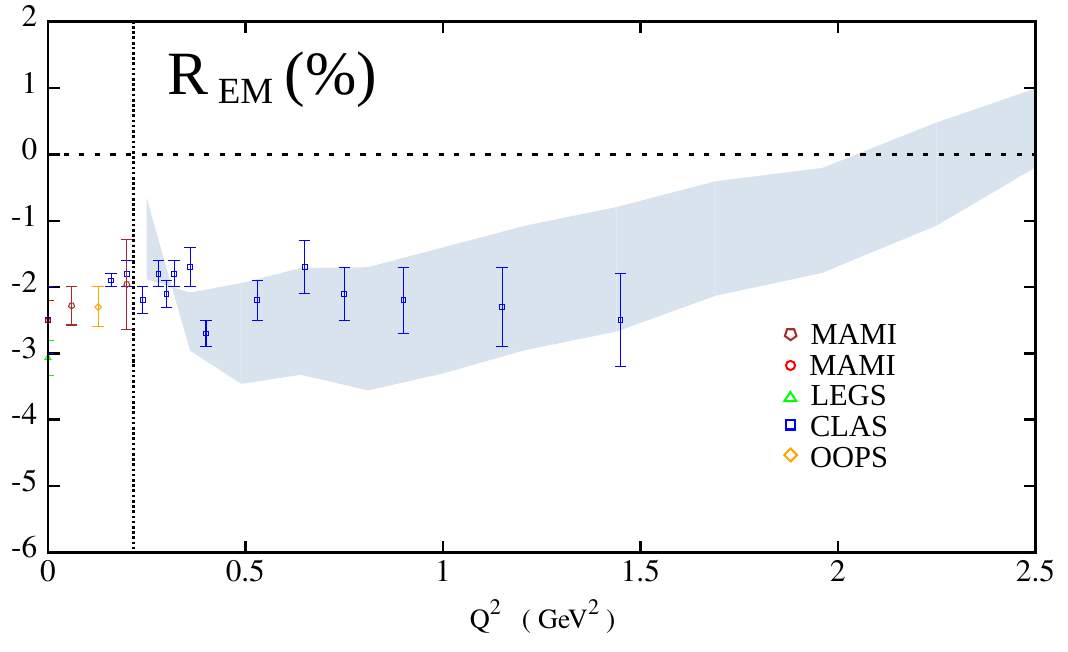}}
\resizebox{0.45\textwidth}{!}{\includegraphics{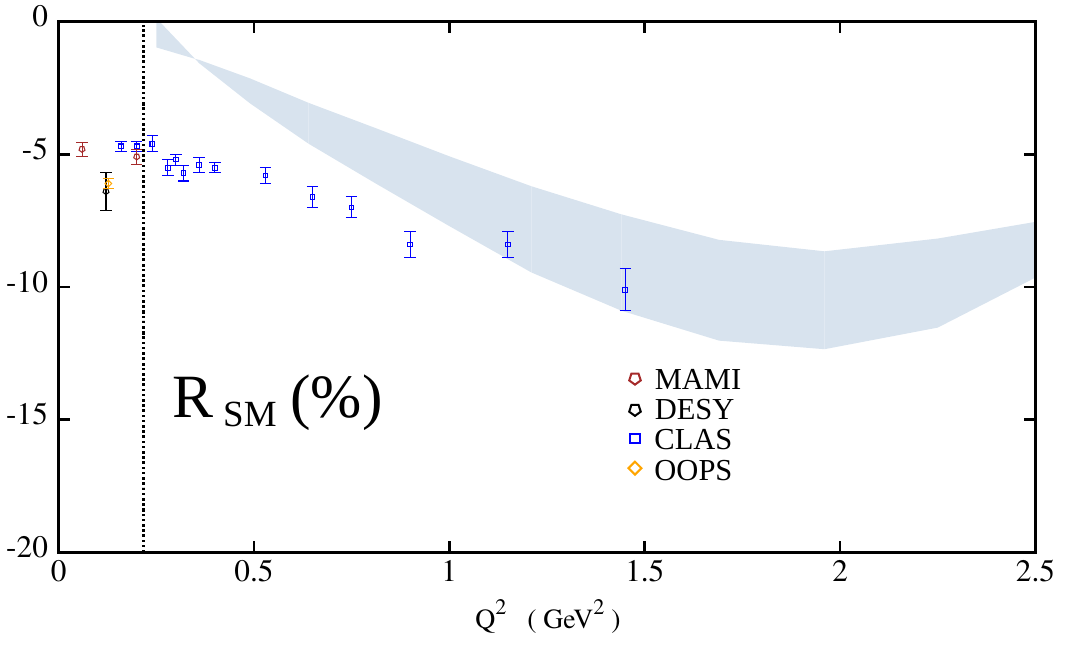}}
\end{center}
\caption{Magnetic form factor $G_M$, ratio $R_{EM}$ and ratio $R_{SM}$ of the 
$\gamma^*N\rightarrow\Delta$ transition.
Vertical dashed line delimit the region below which the singularities of the quark propagator are probed.
Experimental data are taken from 
\cite{Bartel:1968tw,Stein:1975yy,Beck:1999ge,Frolov:1998pw,Pospischil:2000ad,Blanpied:2001ae,Sparveris:2004jn,Stave:2008aa,Aznauryan:2009mx}.
}
\label{fig:NDg}
\end{figure}
We show our results in Fig.~\ref{fig:NDg} compared to experimental data. 
The results for the magnetic form factor are
displayed in the upper panel. 
Within the theoretical systematic uncertainty estimated by the shaded bands in the plots
our results agree
well with the experimental data for $Q^2 > 0.8$ GeV$^2$. Below we observe deviations leading to a systematic underestimation
of the form factor. For the electric and scalar form factors $G_E$ and $G_C$  (not shown) we observe a similar trend, the results agreeing with experiment at moderate to high $Q^2$, and underestimating the data for lower momenta. This trend is in systematic agreement with others for the nucleon electromagnetic and axial form factors as well as the form factors for the $\Delta$ 
\cite{Eichmann:2011vu,Eichmann:2011pv,Sanchis-Alepuz:2013iia}
and indicates, similar as discussed above, the onset of pion cloud effects, 
which are not included in the present framework (see, however,
\cite{Sanchis-Alepuz:2014wea} for pion cloud corrections to baryon masses). 
For the ratio $R_{EM}$, displayed in the
lower left panel of Fig.~\ref{fig:NDg}, we confirm an observation
that has been previously discussed in the quark-diquark approximation \cite{Nicmorus:2010sd}. 
This quantity is sensitive to the presence of deformations of the nucleon and $\Delta$ due to higher angular momentum. In the non-relativistic
quark model, these come into play via $d$-wave admixtures to the $s$-wave structure of the octet baryons. 
In our approach, as discussed above, they are mostly due to the $p$-waves residing in the lower components
of the baryon's spinorial amplitudes.  This observation also extends to the ratio $R_{SM}(0)$, see the
right lower panel of Fig.~\ref{fig:NDg}. 

\subsection*{Hyperon Octet - Decuplet transition}

In the exact $SU(3)$-isospin limit, the transition $\gamma^*\Sigma^{*~+}\rightarrow\Sigma^+$ would be identical to
the $\gamma^*N\rightarrow\Delta$ studied in the shown in the previous subsection. This is indeed
manifest in the magnetic form factor which is comparable in magnitude to the
corresponding one in Fig.~\ref{fig:NDg} and qualitatively identical in shape. Similar remarks apply to
the $\gamma^*\Sigma^{*~0}\rightarrow\Sigma^0$, see ref. \cite{Sanchis-Alepuz:2017mir} for more details.

Once again, in the exact $SU(3)$-isospin limit the $\gamma^*\Xi^{*~0}\rightarrow\Xi^0$ transition would be
identical to the $N\Delta$ or the $\Sigma^{*+}\Sigma^+$ ones. We find that their magnetic form factor is indeed very similar. Since the $\Xi$ is a doubly-strange baryon,
this indicates that the isospin-breaking effects are very small.

A good measure of the breaking of $SU(3)$-flavour symmetry is given by the form factors of the $\gamma^*\Sigma^{*~-}\rightarrow\Sigma^-$ and the $\gamma^*\Xi^{*~-}\rightarrow\Xi^-$ transitions,
since in the limit of exact symmetry these would vanish identically.
Their magnetic form factors indicate a breaking of $SU(3)$ symmetry at the level of a few percent. The smallness of this breaking
in the present calculation might result from the fact that it is only generated by the different quark masses
of the $s-$ and the $u/d-$quarks, both the current quark mass and the dynamically generated one. Other possible
sources of $SU(3)$ breaking would come, for example, from the weakening of the quark-gluon interaction as a
function of the quark mass (see \cite{Williams:2014iea}).
We should mention, however, that such an effect
was explored in \cite{Sanchis-Alepuz:2014sca} in relation to the octet and decuplet baryon masses, where
it was found to be sizeable on the level of the propagators, but extremely small on the level of observables.

\subsection*{$\Sigma^0 - \Lambda$ transition}

The $\gamma^*\Sigma^0\rightarrow\Lambda$ transition is the only electromagnetic transition allowed between members
of the baryon octet. As with other octet hyperon form factors, the experimental information is limited to the magnetic moment \cite{Olive:2016xmw}.
This reaction, however, has attracted considerable theoretical interest. In particular,
the form factors for non-vanishing photon momentum have been studied in \cite{VanCauteren:2005sm,Ramalho:2012ad}. The emphasis of \cite{Ramalho:2012ad} was
on exploring the role of quark-core and meson-cloud effects on the $\gamma^*\Sigma^0\rightarrow\Lambda$ transition
form factors.

\begin{figure}[b]
\begin{center}
\resizebox{0.45\textwidth}{!}{\includegraphics{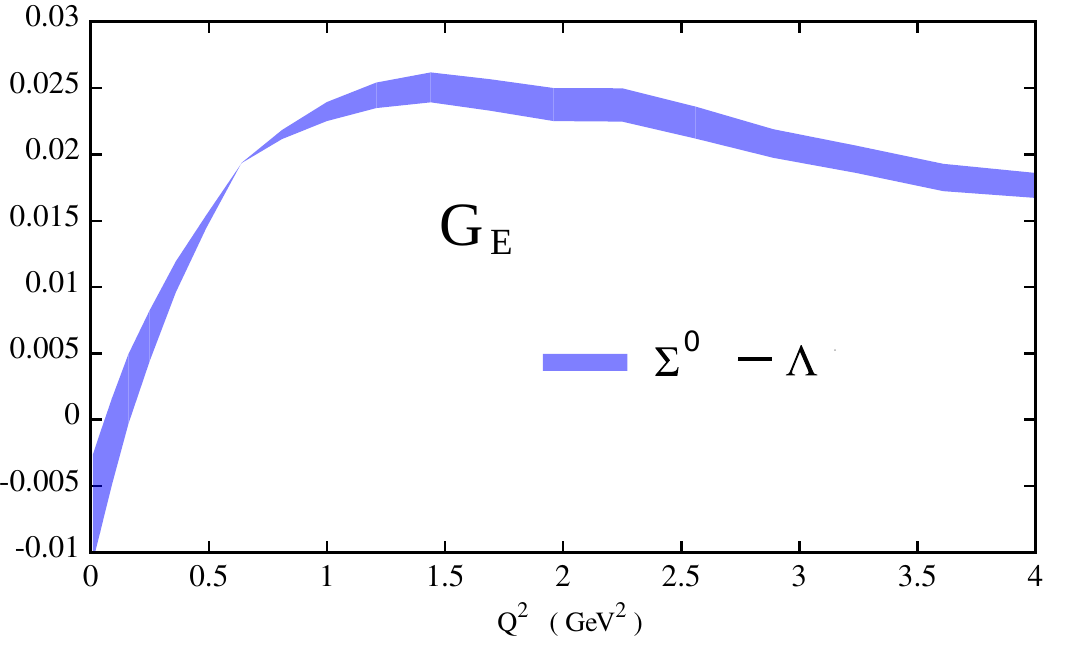}}
\resizebox{0.44\textwidth}{!}{\includegraphics{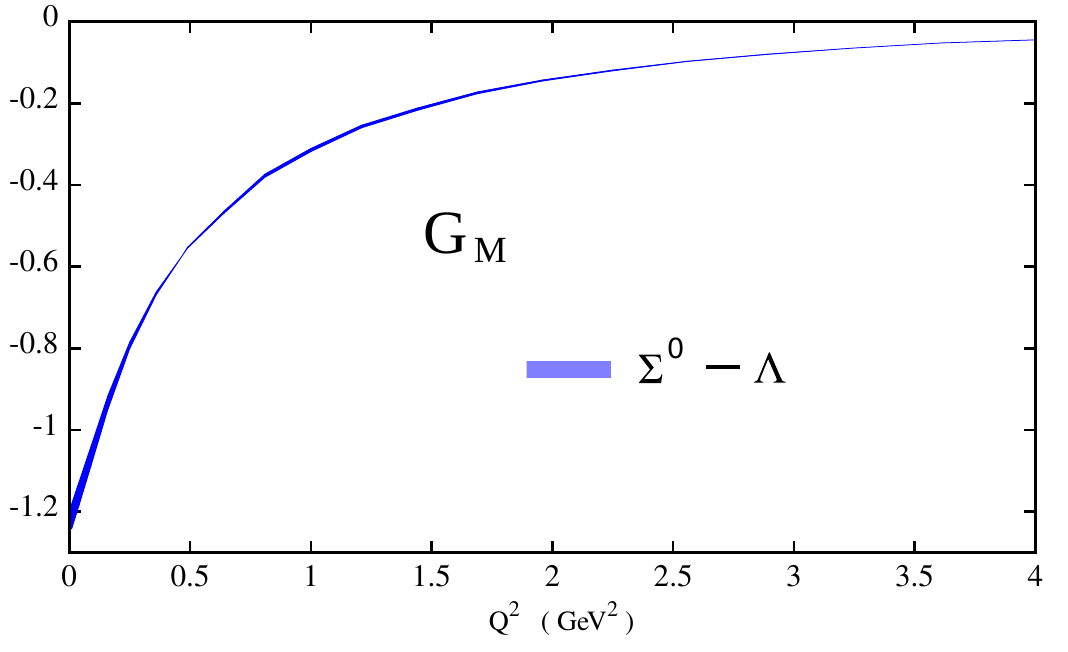}}
\end{center}
\caption{Electric (upper panel) and magnetic (lower panel) form factors of the
octet-only $\gamma^*\Sigma^{0}\rightarrow\Lambda$ transition.
}
\label{fig:sigma0_lambda}
\end{figure}

In contrast to the octet-decuplet transitions, our results in Fig.~\ref{fig:sigma0_lambda} show important
dissimilarities with the quark model results of \cite{Ramalho:2012ad}. There, the quark-core contribution
to the electric form factor is vanishing, being entirely determined by pion-cloud contributions. This is in
contrast to our findings, where we obtain a $G_E$ of similar magnitude to that in \cite{Ramalho:2012ad} but
with meson-cloud effects completely absent. This observation is similar to the one discussed above in connection
with $R_{EM}$ of the nucleon-$\Delta$ transition and the explanation is the same: In quark models with s-wave contributions
to the wave functions only, these transitions are zero by default. Thus one need to invoke either unnaturally
strong d-wave contributions to the wave-function or attribute the entire non-vanishing form factor to meson
cloud effects. In our relativistic framework, however, sizeable p-wave contributions to the baryons wave function
appear naturally and thus generate a non-zero result for $G_E$.

\section{Conclusions and Outlook}

The here presented results are obtained in the so-called rainbow-ladder truncation with a widely 
used form of the quark-gluon interaction, the Maris-Tandy model \cite{Maris:1999nt}.
It allowed for a unified and highly successful description of mesons and baryons, see, {\it e.g.},
refs.\ \cite{Eichmann:2016yit,Bashir:2012fs} and references therein. The somewhat surprising
success of the rainbow-ladder truncation as well as its limitations have been discussed in great 
detail in the literature, see, {\it e.g.}, refs.\ \cite{Sanchis-Alepuz:2015qra,Sanchis-Alepuz:2015tha} 
for a recent synopsis. A further and important novel development is the calculation of
the spectrum of light mesons 
in a symmetry-preserving truncation based on the three-particle irreducible effective action
\cite{Williams:2015cvx}. The obtained results hereby not only verify  the previous arguments 
on the reliability and the limitations of the rainbow-ladder truncation but mark a very decisive 
step towards the goal of apparent convergence in the pattern of increasingly sophisticated 
truncations within the Dyson-Schwinger--Bethe-Salpeter approach. An extension of calculations
within this truncation to baryons is in progress, and it will be interesting to see whether and 
how the obtained results will deepen our understanding of structure of baryons.

\section*{Acknowledgements} 
R.A. is grateful to the organizers of the 
International Conference on Exotic Atoms and Related Topics - EXA2017 
for all their efforts which made this conference possible, and for financial support. 

This work has been supported by the project P29216-N36
from the Austrian Science Fund, FWF, by the Helmholtz International Center
for FAIR within the LOEWE program of the State of Hesse, and by the
DFG grant FI 970/11-1.

\bigskip \bigskip

\bibliography{EXA17Alkofer}

\end{document}